\documentclass[a4paper]{spie}  
\usepackage[]{graphicx}

\title{ALMA Temporal Phase Stability and the Effectiveness of
	Water Vapor Radiometer}

\author{Satoki Matsushita\supit{a,b}, Koh-Ichiro Morita\supit{c,b},
	Denis Barkats\supit{b}, Richard E. Hills\supit{b},
	Ed Fomalont\supit{d}, and Bojan Nikolic\supit{e}
\skiplinehalf
\supit{a}Academia Sinica Institute of Astronomy and Astrophysics,
	P.O.\ Box 23-141, Taipei 10617, Taiwan, R.O.C.; \\
\supit{b}Joint ALMA Observatory, Alonso de C\'ordova 3107, Vitacura
	763 0355, Santiago, Chile; \\
\supit{c}National Astronomical Observatory of Japan, 2-21-1 Osawa,
	Mitaka, Tokyo 181-8588, Japan; \\
\supit{d}National Radio Astronomy Observatory, Charlottesville,
	VA 22903, USA; \\
\supit{e}Astrophysics Group, Cavendish Laboratory, University of
	Cambridge, J J Thomson Avenue, Cambridge CB3 0HE, United Kingdom
}

\authorinfo{Further author information: (Send correspondence to S.M.)\\
	S.M.: E-mail: satoki@asiaa.sinica.edu.tw, Telephone: 886 (0)2 3365 2200}

\begin{document} 
  \maketitle 

\begin{abstract}
Atacama Large Millimeter/submillimeter Array (ALMA) will be the
world largest mm/submm interferometer, and currently the Early
Science is ongoing, together with the commissioning and science
verification (CSV).
Here we present the temporal phase stability of the entire ALMA
system; from antenna to correlator.
The data, taken during the last 2 years of CSV activities, consisted
of integrations on strong point sources (i.e., bright quasars) at
various frequency bands, and at various baseline lengths (up to
600 m).
We verified the temporal phase stability of ALMA using these data.
We observed a strong quasar for a long time (from a few tens of
minutes, up to an hour), derived the 2-point Allan Standard Deviation
after the atmospheric phase correction using the 183 GHz Water Vapor
Radiometer (WVR) installed in each 12 m antenna, and confirmed that
the phase stability of all the baselines reached the ALMA
specification.
Since we applied the WVR phase correction to all the data mentioned
above, we also studied the effectiveness of the WVR phase correction
at various frequencies, baseline lengths, and weather conditions.
The phase stability often improves a factor of 2 - 3 after the
correction, and sometimes a factor of 7 improvement can be obtained.
However, the corrected data still displays an increasing phase
fluctuation as a function of baseline length, suggesting that the dry
component (e.g., N$_{2}$ and O$_{2}$) in the atmosphere also
contributes the phase fluctuation in the data, although the
imperfection of the WVR phase correction cannot be discarded at this
moment.
\end{abstract}



\section{INTRODUCTION}
\label{sect-intro}  

Atacama Large Millimeter/submillimeter Array (ALMA)\cite{hil10} will
be the world largest millimeter-/submillimeter-wave interferometer,
and currently the Early Science observations are ongoing, together
with the Commissioning and Science Verification (CSV).
One of the important tasks for CSV is to verify that the stabilities
of the whole ALMA system are satisfying the specifications.
Within this stability verification, verifying phase stability is one
of the key missions, since the interferometric phase has the position
and distribution information of the emission from target astronomical
sources, and therefore critical for imaging the astronomical sources
with high dynamic range and accurate positions.

The ALMA specification for the long-term temporal phase stability is
defined as follows\cite{sra06}:
The total instrumental phase error, $\sigma$, should be below
35 femtosecond per baseline (i.e., 25 femtosecond per antenna) that
is calculated with the 2-point Allan Standard Deviation (hereafter
2-p ASD),
\begin{equation}
\sigma = \sqrt{\frac{<\{\phi_\tau(t+T) - \phi_\tau(t)\}^2>}{2}},
\end{equation}
where $\phi_{\tau}$ is phase with a fixed averaging time, $\tau$, of
10 second and intervals, $T$, between 20 and 300 second.
$t$ indicates the data sample time, and $<...>$ means the average
over the data sample.
This equation is almost the same as the definition of the temporal
structure function for the phase fluctuation\cite{tat61,hol95,tho01},
except the temporal smoothing of 10 seconds, to measure the long-term
characteristics with avoiding the short-term noise, such as
high-frequency atmospheric/instrumental noise and thermal noise from
various instruments.

In this proceeding, we show the ALMA system verification results of
temporal phase stability.
For this verification, it is crucial to eliminate all the
instabilities that originate outside the ALMA system, and the
dominant instability is that originates from water vapor in the
atmosphere.
To eliminate the instability caused by water vapor, ALMA is adopting
a phase correction method using the 183 GHz Water Vapor Radiometer
(WVR)\cite{sti04,nik09a,nik09b}.
In this proceeding, we also provide the impact of the WVR phase
correction to the ALMA data.

\section{MEASUREMENTS AND DATA ANALYSIS}
\label{sect-meas}  

To measure the long-term temporal phase stability of the whole ALMA
system, we observed strong point sources, namely radio loud quasars,
for $\sim20-60$ minutes in the ALMA bands 3 ($\sim86$ GHz),
6 ($\sim230$ GHz), 7 ($\sim345$ GHz), and 9 ($\sim690$ GHz) using
$6-16$ twelve meter diameter antennas under various weather
conditions, which are the precipitable water vapor (PWV) of
$0.2-2.9$ mm.
The details of the measurements are listed in Table~\ref{tab-meas}.

%

\begin{table}[t]
\caption{Details of the measurements.
	(1) Data identification number.
	(2) ALMA band name (number).
    (3) Measurement frequency in GHz.
    (4) Measurment date.
	(5) Precipitable water vapor in mm.
    (6) Observed source name.
    (7) Total observation time in minute.
    (8) Integration time of each data point in second.
    (9) Number of antennas used in the measurement.
	}
\label{tab-meas}
\begin{center}
\begin{tabular}{|c|cc|cccccc|}
\hline
\rule[-1ex]{0pt}{3.5ex}
 Data No. & Band & Freq. &     Date     & PWV  &   Source   & Obs.~Time & Int.~Time & Ant.~No. \\
          &      & [GHz] & [YYYY/MM/DD] & [mm] &            &   [min]   &   [sec]   & \\
\hline
 1 & 3 &  87.2 &  2010/10/26  & 0.3  & $0538-440$ &    60     &    0.1    &  7 \\
 2 &   &  86.3 &  2010/11/08  & 0.5  & $0538-440$ &    60     &    0.1    &  7 \\
\hline
 3 & 6 & 229.5 &  2010/10/23  & 0.7  & $0538-440$ &    60     &    1.0    &  8 \\
 4 &   &       &              & 0.6  & $0538-440$ &    60     &    0.5    &  8 \\
 5 &   &       &  2010/10/26  & 0.3  & $0538-440$ &    60     &    0.1    &  7 \\
 6 &   & 230.6 &  2010/12/16  & 0.6  & $0538-440$ &    60     &    1.0    &  7 \\
 7 &   & 229.5 &  2010/12/21  & 2.9  &    3C279   &    60     &    1.0    &  6 \\
\hline
 8 & 7 & 344.8 &  2010/10/26  & 0.3  & $0538-440$ &    60     &    0.1    &  7 \\
 9 &   & 333.9 &  2011/10/06  & 0.2  & $0522-364$ &    20     &    1.0    & 14 \\
10 &   &       &              & 0.2  & $0522-364$ &    20     &    1.0    & 14 \\
11 &   &       &  2011/10/09  & 1.1  & $0538-440$ &    20     &    1.0    & 13 \\
\hline
12 & 9 & 687.7 &  2011/09/11  & 0.2  & $1924-292$ &    20     &    1.0    & 13 \\
13 &   &       &  2011/10/08  & 0.5  & $0522-364$ &    20     &    1.0    & 16 \\
14 &   &       &  2012/03/21  & 0.8  &    3C279   &    20     &    1.0    &  7 \\
\hline
\end{tabular}
\end{center}
\end{table} 

The WVR phase correction has first been applied to the obtained data
sets using the {\tt wvrgcal} program. 
The temporal phase stability of each data set has then been
calculated using a personally developed python program under the
Common Astronomy Software Applications (CASA) package environment.
For the 10 second data binning, if the data points are less than 70\%
of the number of the data points should be, this binned data point
has been discarded (for example, assume the integration time of one
data point is 1 second, and if the data points are less than 7 data
points within 10 second, then this binned 10 second averaged data
point has not been used).
Phase error in femtosecond has been calculated as
[rms phase in degree] / $360^{\circ}$ / [measurement frequency in Hz].


\section{Effectiveness of the Water Vapor Radiometer Phase Correction}
\label{sect-wvr}

\begin{figure}
\begin{center}
\begin{tabular}{c}
\includegraphics[width=16cm]{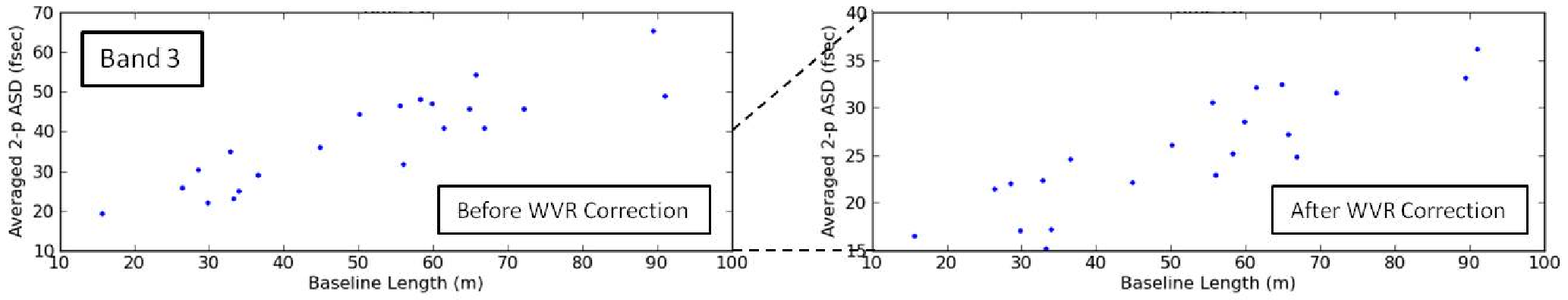}\\
\includegraphics[width=16cm]{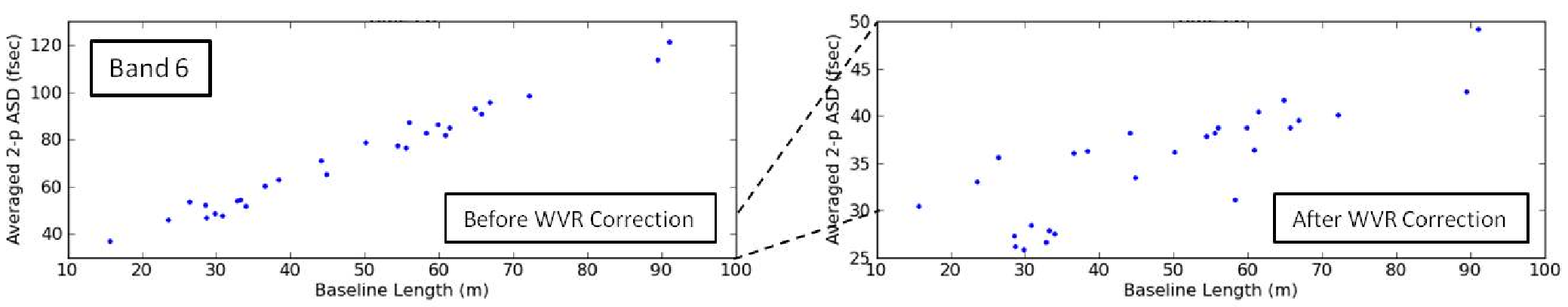}\\
\includegraphics[width=16cm]{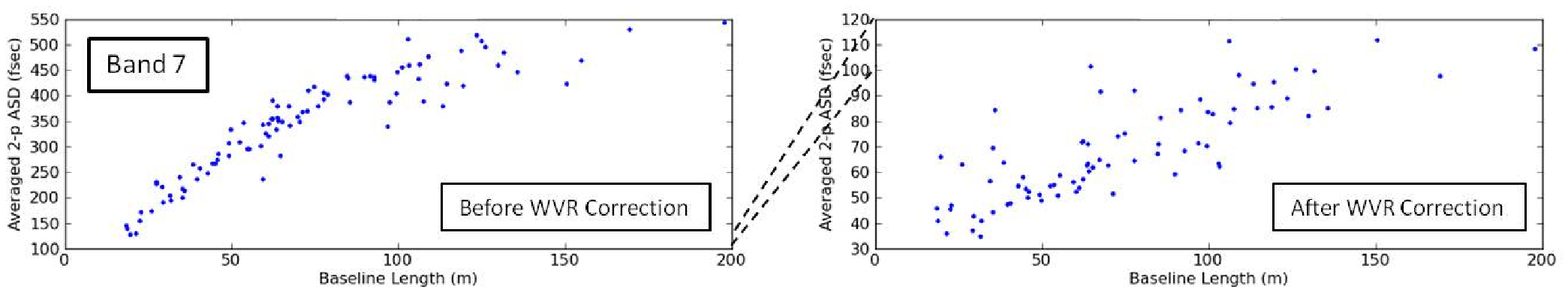}\\
\includegraphics[width=16cm]{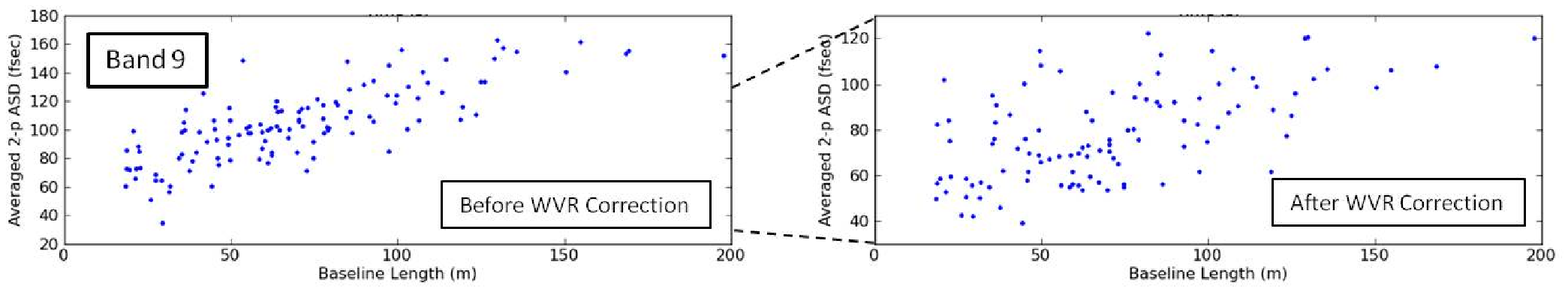}
\end{tabular}
\end{center}
\caption{
	\label{fig-WVRimproved}
	SSF plots before (left plots) and after (right plots) the WVR
	phase correction in all four ALMA bands.
	Here, we exhibit the results, which significantly improved the
	phase stability due to the WVR phase correction.
	Since the scaling of the vertical axes is different between the
	plots before and after, the same range is indicated with dashed
	lines.
	({\it 1st row}) Band 3 (data No.~2) case under PWV of 0.5 mm.
	({\it 2nd row}) Band 6 (data No.~3) case under PWV of 0.7 mm.
	({\it 3rd row}) Band 7 (data No.~11) case under PWV of 1.1 mm.
	({\it 4th row}) Band 9 (data No.~13) case under PWV of 0.5 mm.
	}
\end{figure}

First, we present the effectiveness of the WVR phase correction to
the measurement data sets in all the ALMA bands.
To show this, we use spatial structure function (structure function
as a function of baseline length; hereafter SSF), which is, average
the rms phase in the 2-p ASD of interval $T$ between 20 and 300
second for each baseline, and plot this as a function of baseline
length.
Fig.~\ref{fig-WVRimproved} displays some example SSF plots that
improved significantly by the WVR phase correction.
The improvement of the phase fluctuation can be seen from short
($\sim10-20$ m) to long (up to 200 m) baseline lengths, with a factor
of $\sim2$ improvement or even more; a factor of $\sim4$ for the
band 7 case in Fig.~\ref{fig-WVRimproved}.
The data that exhibit significant improvement are often taken under
between normal to good weather conditions; the examples shown in
Fig.~\ref{fig-WVRimproved} were taken under the PWV of
$0.5\sim1.1$ mm.

\begin{figure}
\begin{center}
\begin{tabular}{c}
\includegraphics[width=8cm]{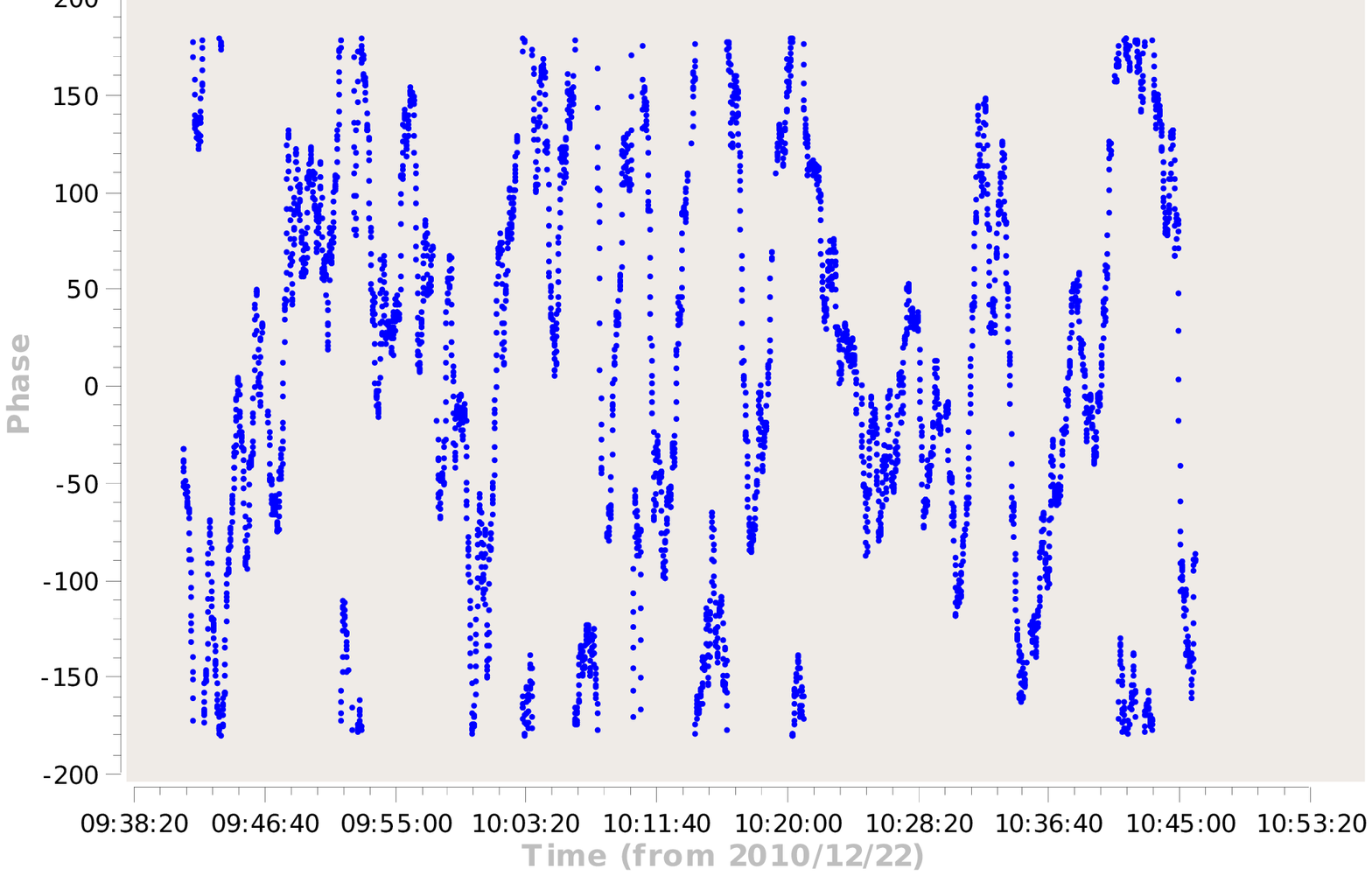}
\includegraphics[width=8cm]{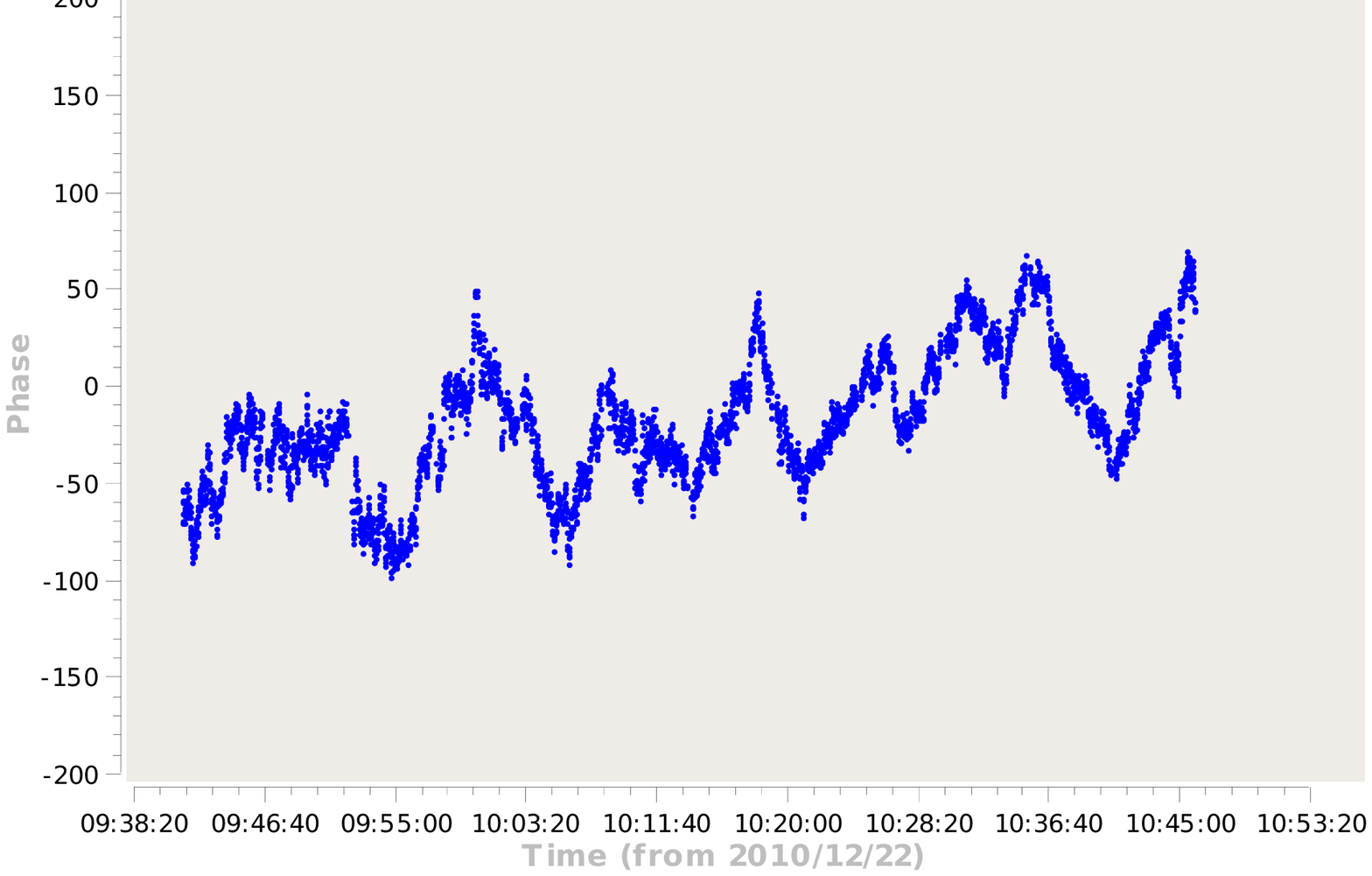}
\end{tabular}
\end{center}
\vspace{-6cm}
\caption{
	\label{fig-wvr600m}
	Time variation plots of phase before (left plot) and after (right
	plot) the WVR phase correction under the PWV of 2.9 mm and the
	baseline length of $\sim600$ m.
	It can be clearly seen that the degree of phase fluctuation
	significantly improved after the WVR phase correction.
	}
\end{figure}

The effectiveness of the WVR phase correction can also be seen in
much longer baseline length of $\sim600$ m.
Fig.~\ref{fig-wvr600m} displays the phase fluctuation plot as a
function of time before and after the WVR phase correction using the
data No.7 (band 6 data).
The raw data (before the WVR phase correction) shows large phase
fluctuation that is more than $\pm180^{\circ}$
(Fig.~\ref{fig-wvr600m} left plot), namely if we integrate this data
set, the resultant data will lose coherence, but after the WVR phase
correction, the phase fluctuation kept within $\pm50^{\circ}$
(peak-to-peak; a factor of $\sim4$ improvement) with a long-timescale
phase drift (Fig.~\ref{fig-wvr600m} right plot), which is good
quality enough to use for the astronomical data analysis.
It is important to mention that the weather condition was not good
when this band 6 long baseline length data set has been taken (PWV
$\sim2.9$ mm), but this result prove that the data can be useful even
it is taken under not really good weather condition, once the WVR
phase correction has been applied.

\begin{figure}
\begin{center}
\begin{tabular}{c}
\includegraphics[width=16cm]{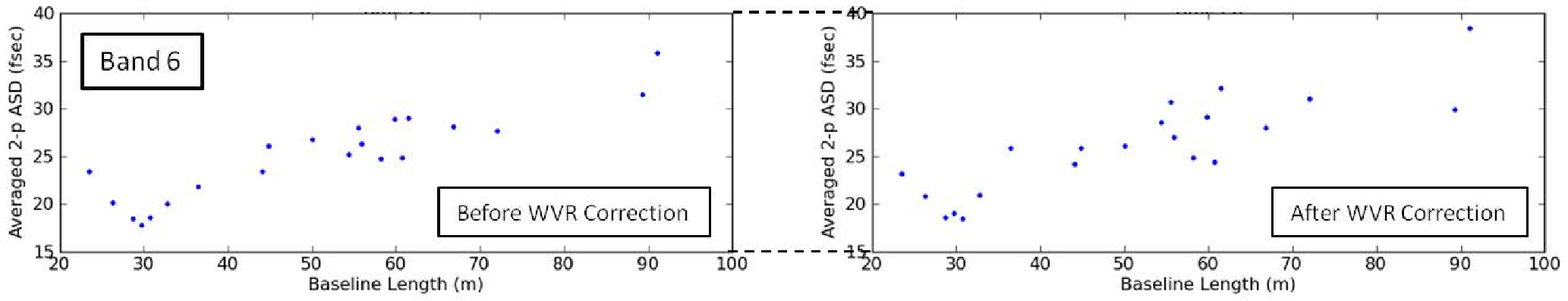}\\
\includegraphics[width=16cm]{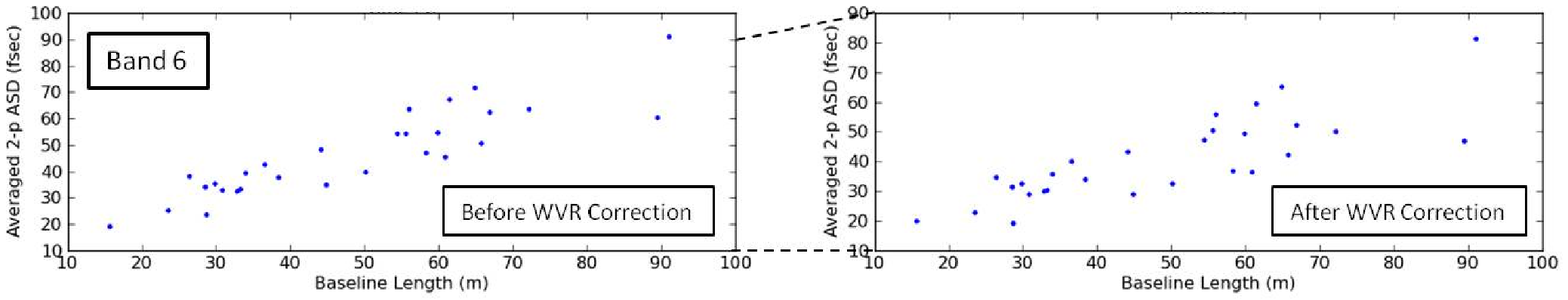}
\end{tabular}
\end{center}
\caption{
	\label{fig-WVRnotimproved}
	SSF plots before (left plots) and after (right plots) the WVR
	phase correction in band 6.
	Here, we show the results that do not improved the phase
	stability with the WVR phase correction.
	({\it 1st row}) Band 6 (data No.~5) case under PWV of 0.3 mm.
	({\it 2nd row}) Band 6 (data No.~4) case under PWV of 0.6 mm.
	}
\end{figure}

On the other hand, some data did not improve much
(Fig.~\ref{fig-WVRnotimproved}).
The data that do not show the improvement are often taken under very
good weather conditions.
Data No.~1, 5, 9, and 12 are the examples for this case (only the
band 6 plots are shown in the 1st row of
Fig.~\ref{fig-WVRnotimproved}); these data are taken under the PWVs
of $0.2\sim0.3$ mm.
This is understandable, since the phase fluctuation is mainly caused
by water vapor in the atmosphere, and under low PWV condition, the
phase fluctuation itself is already low.
It is clearly seen in the maximum values of the vertical axis (SSF in
femtosecond, which corresponds to the rms phase fluctuation) of
Fig.~\ref{fig-WVRnotimproved} compared with those in the previous
plots in Fig.~\ref{fig-WVRimproved}.
Under this low PWV condition, the difference in fluctuations of PWVs
between the line of sights of two antennas (and therefore the
difference in fluctuations of two WVR outputs) is too low, and
therefore not much improvement has been seen by the WVR phase
correction.
In addition, phase fluctuation due to the dry component in the
atmosphere (see the following two paragraphs below) may dominate the
atmospheric phase fluctuation, and therefore resulted as less phase
correction effect by WVR.
%
But in some data set with low chance of occurrence, the phase
fluctuation did not improve even under worse PWV condition of 0.6 mm;
the 2nd row of Fig.~\ref{fig-WVRnotimproved} displays an example of
this case, probably dominated either by the dry or ice component in
the atmosphere.
From these examples, the WVR phase correction does not work well
(or in other word, do not need to do) under very good weather
conditions (PWV $\leq0.3$ mm), but often works under worse conditions
(PWV $\sim0.5-2.9$ mm).
But in some rare case, it will not work even under worse conditions.

An interesting point we should note is that even under very good
weather conditions or after the WVR phase correction, the rms phase
fluctuation increases as a function of baseline length
(Figs.~\ref{fig-WVRimproved}, \ref{fig-WVRnotimproved}).
If something caused by instruments, problems will show up as the
antenna-base, not as the baseline-base, so it will not show up as a
function of baseline length.
For example, ALMA has the line length corrector that compensate the
signal path (and therefore the phase) change of the cables for each
antenna, and a problem in the line length corrector result as an
antenna-base problem.
This suggests that there is another component other than water vapor
that cause the phase fluctuation exists in the atmosphere.
The most probable component is the dry component (N$_{2}$ and/or
O$_{2}$) in the atmosphere.
Phase fluctuation due to the dry component is believed to be caused
by changes in density of the dry air, and the phase fluctuation on
short baselines is most likely due to the changes in temperature.
The temporal phase structure function of the dry component will tell
the temporal density/temperature fluctuation of the dry air, and this
will be an important study for the coming long baseline operation of
ALMA.
The existence of the dry component in the phase fluctuation means
that there is a limit to correct phase fluctuation only with WVR,
and some other methods are needed to correct phase fluctuation
perfectly.

On the other hand, there may be other possible causes of this trend,
such as the WVR phase correction is still not optimized and therefore
the residual of the phase correction appears more in longer
baselines, or the WVR phase correction reached to the performance
limit.
However, in case of performance limit, it should show an upper limit
in the temporal phase structure function at long baseline, since
longer the baseline length, more difference in the atmospheric water
vapor content, and eventually the difference of the water vapor
contents between two antennas surpass the performance limit of the
WVR phase correction.
Future long baseline measurements will make this point clear.

\section{Temporal Phase Stability and the Verification of the ALMA
	Specification}
\label{sect-stab}

\begin{figure}
\begin{center}
\begin{tabular}{c}
\includegraphics[width=15cm]{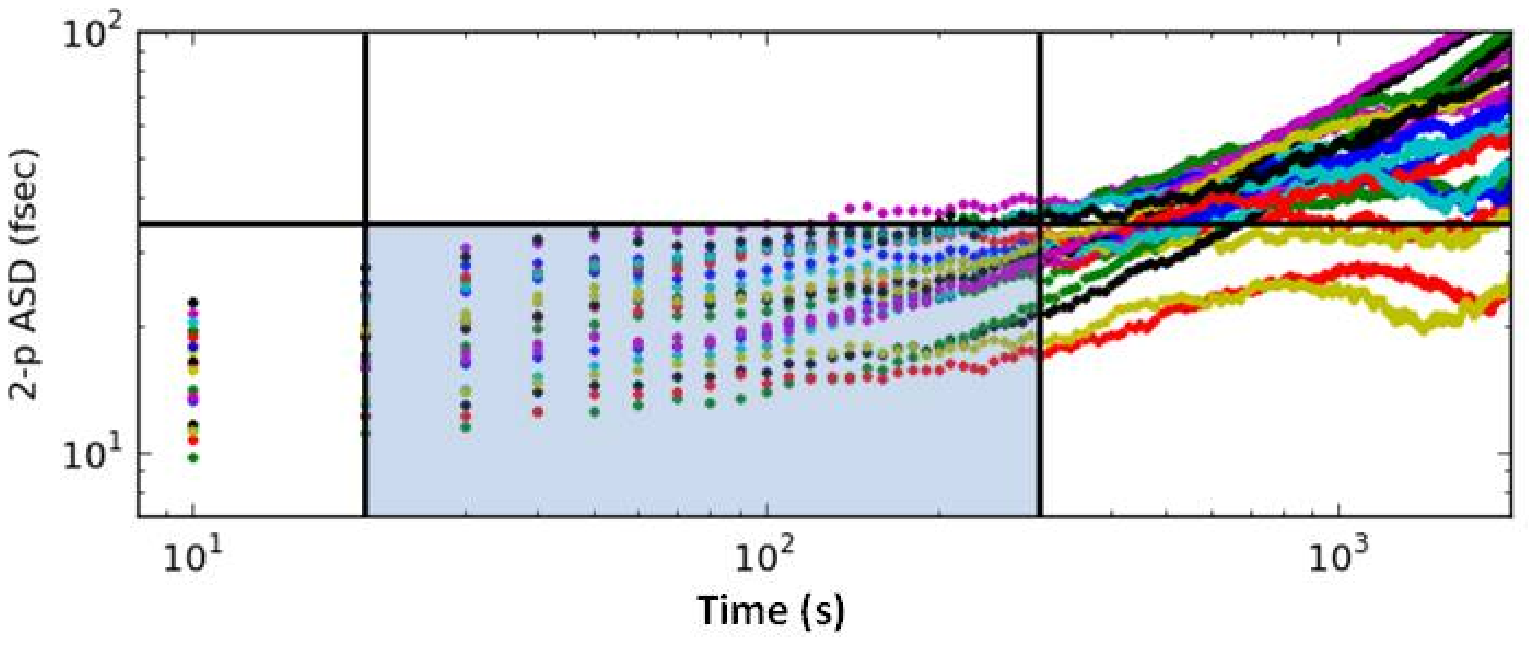}\\
\includegraphics[width=15cm]{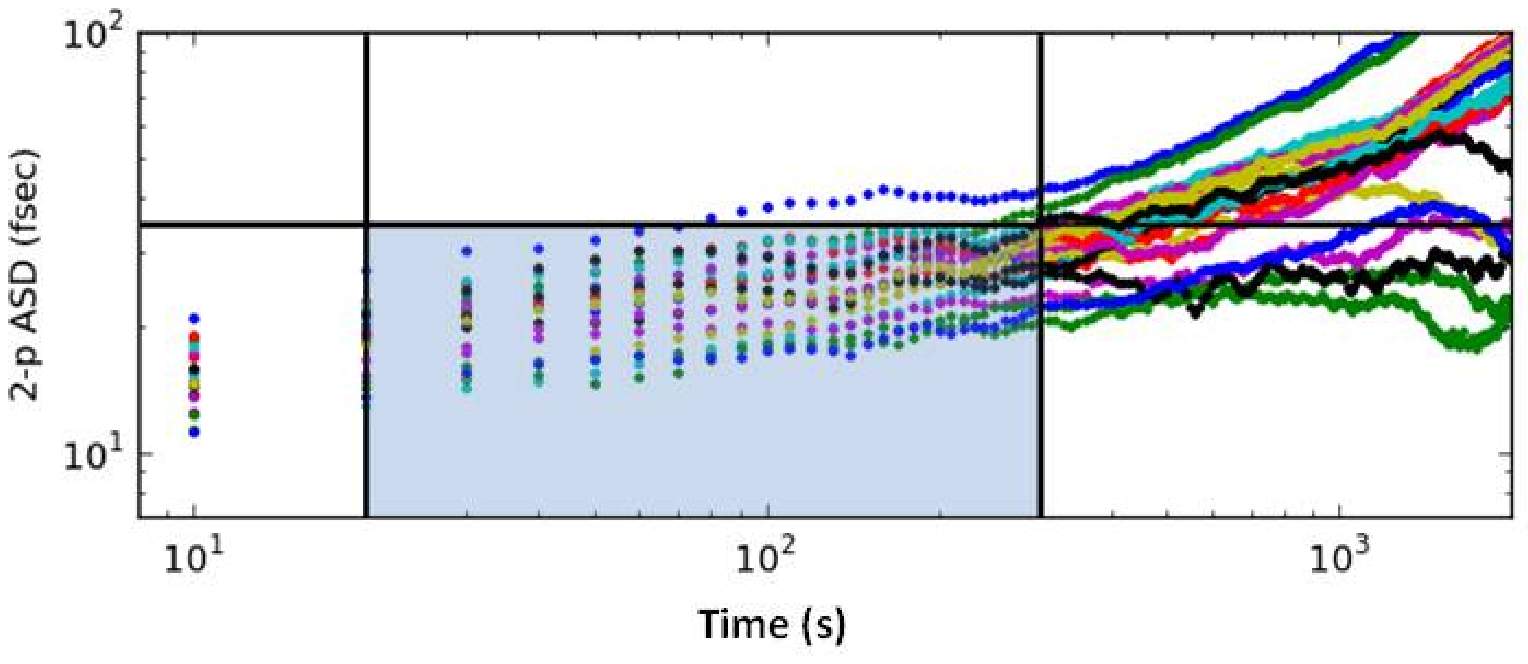}\\
\includegraphics[width=15cm]{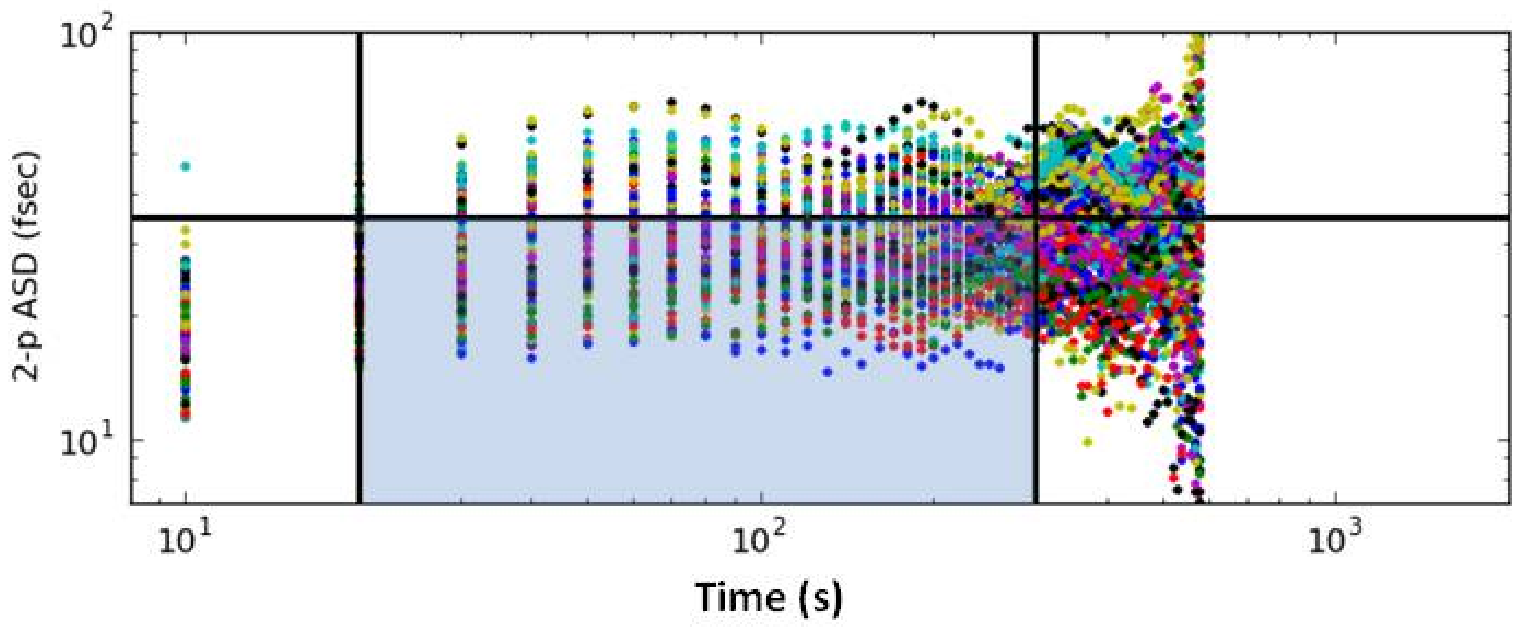}
\end{tabular}
\end{center}
\caption{
	\label{fig-2pASD}
	2-p ASD plots that reach the specification in ({\it 1st row})
	band 3, ({\it 2nd row}) band 6, and ({\it 3rd row}) band 7.
	}
\end{figure}

After taking out (most of) the atmospheric effect, it is possible to
verify whether the ALMA system has reached to the specification or
not.
Since SSF increases as a function of baseline length even after
the phase correction, we should judge whether the ALMA temporal phase
stability reaches the specification or not at the short baseline
region.

Fig.~\ref{fig-2pASD} displays the 2-p ASD plots for bands 3, 6, and
7.
The shaded area is the ALMA specification range.
In these three bands, some or most of the baselines are within the
ALMA specification range.
As mentioned above, some of the baselines that are not in the ALMA
specification range are long baseline length data, which we think
those are affected by the dry air component, so we do not need to
consider here.
We therefore can conclude that the ALMA system for bands 3, 6, and 7
satisfies the ALMA specification.

All the band 9 data we took, on the other hand, did not reach the
ALMA specification, mainly due to low signal-to-noise (S/N) ratio of
the data.
In band 9, the number of bright sources with compact structure is
extremely limited; all the planets are too big for band 9 even with
short baseline lengths, and the minor planets are generally too weak
for this purpose.
Furthermore, weather conditions for band 9 are very severe, and a
slight degrade of weather condition makes the data quality
significantly worse.
On the other hand, the current software only uses the data of
2 GHz bandwidth and one polarization.
If the software has been revised to use full bandwidth (8 GHz) with
two polarization, the sensitivity increases for $2\sqrt{2}$, which
will increase S/N, and this will be the next important update.

\acknowledgments     

We deeply regret the loss of our colleague, Koh-Ichiro Morita, who
lost his life on May 7th, 2012, at Santiago, Chile.
He significantly helped not only this work but also for various works
in the ALMA project, and without him, we could not achieve these
results.
Furthermore, he taught S.M.\ the basic and application of
interferometry, phase correction methods, and site testings since
S.M.\ was a graduate student at the Nobeyama Radio Observatory, and
without his guidance S.M.\ could not be at the current position.
Morita-san, we really miss you...

S.M.\ is supported by the National Science Council (NSC) of Taiwan,
NSC 100-2112-M-001-006-MY3.


\end{document}